\documentclass[sigconf]{acmart}
\AtBeginDocument{%
  \providecommand\BibTeX{{%
    \normalfont B\kern-0.5em{\scshape i\kern-0.25em b}\kern-0.8em\TeX}}}

\copyrightyear{2021}
\acmYear{2021}
\setcopyright{rightsretained}
\acmConference[SIGIR '21]{Proceedings of the 44th International ACM SIGIR Conference on Research and Development in Information Retrieval}{July 11--15, 2021}{Virtual Event, Canada}
\acmBooktitle{Proceedings of the 44th International ACM SIGIR Conference on Research and Development in Information Retrieval (SIGIR '21), July 11--15, 2021, Virtual Event, Canada}\acmDOI{10.1145/3404835.3463253}
\acmISBN{978-1-4503-8037-9/21/07}

\acmConference[SIGIR '21]{SIGIR '21: 44th International ACM SIGIR Conference
on Research and Development in Information Retrieval}{June, 2021}{Online}
\acmBooktitle{SIGIR '21: The 44th International ACM SIGIR Conference
on Research and Development in Information Retrieval, June 2021}
\usepackage[hide]{chato-notes} 
\usepackage{multirow}

\usepackage{pgfplots}
\usepackage{tablefootnote}
\usepackage{subcaption}
\usepackage[utf8]{inputenc}
\usepackage{pgfplots}
\DeclareUnicodeCharacter{2212}{−}
\usepgfplotslibrary{groupplots,dateplot}
\usetikzlibrary{patterns,shapes.arrows}
\pgfplotsset{compat=newest}

\begin{document}
\fancyhead{} 
\title{Wiki-Reliability: A Large Scale Dataset for Content Reliability on Wikipedia}


\author{KayYen Wong}
\authornote{This work was done during KayYen's internship at the Wikimedia Foundation.}

\affiliation{%
  \institution{Outreachy}
  \city{Kuala Lumpur}
  \country{Malaysia}}
\email{kayyenwong@gmail.com}

\author{Miriam Redi}
\affiliation{%
  \institution{Wikimedia Foundation}
  \city{London}
  \country{United Kingdom}}  
  \email{miriam@wikimedia.org}

\author{Diego Saez-Trumper}
\affiliation{%
 \institution{Wikimedia Foundation}
 \city{Barcelona}
 \country{Spain}}
   \email{diego@wikimedia.org}


\begin{abstract}
Wikipedia is the largest online encyclopedia, used by algorithms and web users as  a central hub of reliable information on the web. The quality and reliability of Wikipedia content is maintained by a community of volunteer editors. 
 Machine learning and information retrieval algorithms could 
help scale up editors' manual efforts around Wikipedia content reliability. However, there is a lack of large-scale data to support the development of such research. To fill this gap, in this paper, we propose \textit{Wiki-Reliability}, the first dataset of English Wikipedia articles annotated with a wide set of content reliability issues. 

To build this dataset, we rely on Wikipedia ``templates''.  Templates are tags used by expert Wikipedia editors to indicate content issues, such as the presence of “non-neutral point of view” or “contradictory articles”, and serve as a strong signal for detecting reliability issues  in a revision. 
We select the 10 most popular reliability-related templates on Wikipedia, and 
propose an effective method to label almost 1M samples of Wikipedia article revisions as \textit{positive} or \textit{negative} with respect to each template. Each positive/negative example in the dataset comes with the full article text and 20 features from the revision's metadata. We provide an overview of the possible downstream tasks enabled by such data, and show that \textit{Wiki-Reliability} can be used to train large-scale models for content reliability prediction.
We release all data and code for public use.
\end{abstract}

\ccsdesc[500]{Information systems~World Wide Web}
\ccsdesc[500]{Information systems~Information retrieval}
\ccsdesc[500]{Computing methodologies~Machine learning}

\keywords{Wikipedia, Dataset, NLP, Reliability, Hoaxes, POV, Unreferenced}


\maketitle

\section{Introduction}
Wikipedia is one the largest and most widely used knowledge repositories in the world. People use Wikipedia for studying, fact checking and a wide set of different information needs \cite{singer2017we}.  In developing countries, Wikipedia is largely used as an educational resource \cite{lemmerich2019world}. Moreover, not only people, but AI-based systems learn from Wikipedia \cite{mikolov2017advances} and use it as ground-truth for fact checking \cite{thorne2018fever}. Therefore, the quality of content on Wikipedia is relevant to both humans and machines, and more broadly for the integrity of all human knowledge.

\begin{figure}
    \centering
    \includegraphics[width=0.95\linewidth]{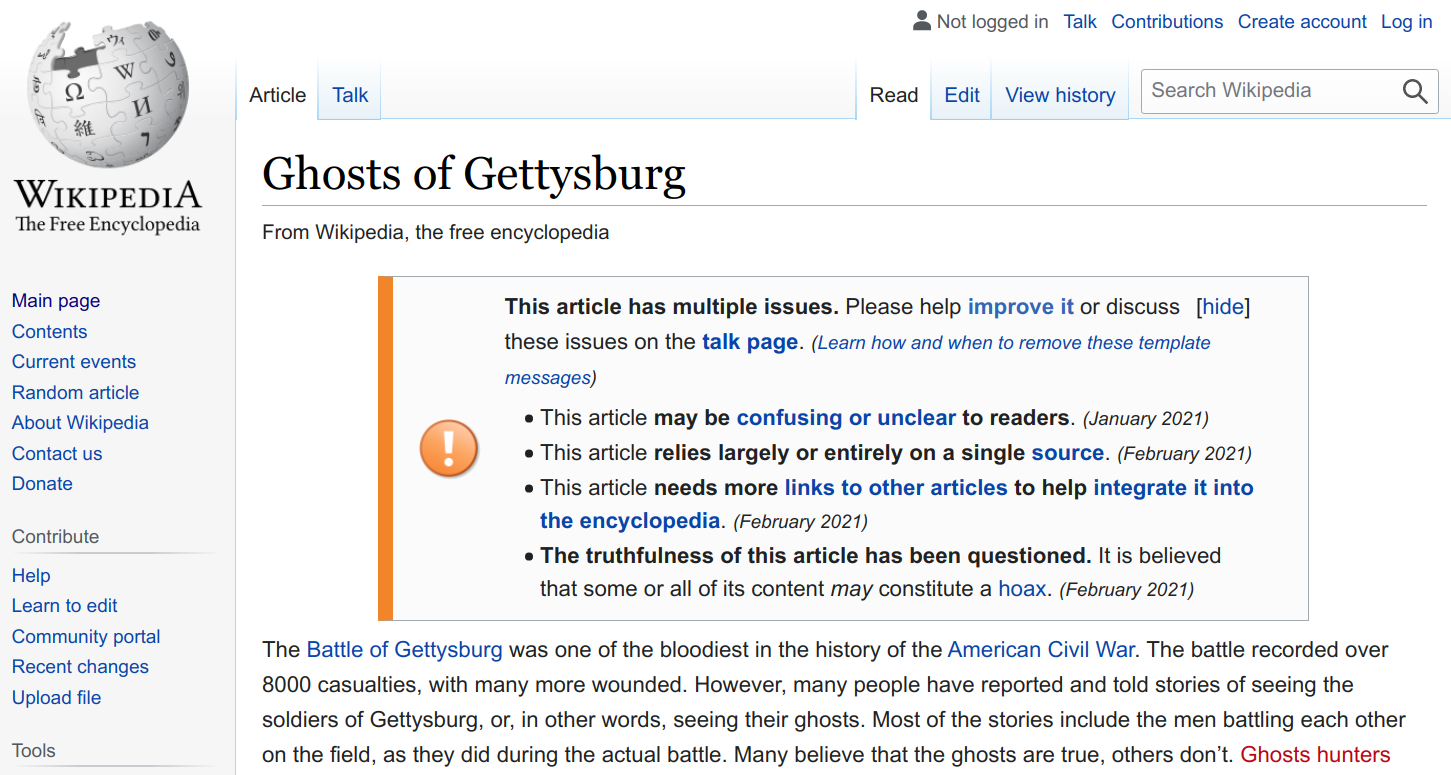}
    \caption{Example of an English Wikipedia page with several template messages describing reliability issues.}
    \label{fig:teaser}
\end{figure}
On Wikipedia, the review and moderation of content is self-governed by Wikipedia’s volunteer community of editors, through collaboratively created policies and guidelines \cite{Beschastnikh08, Forte09}. Despite the large size of the Wikipedia editors community (41k monthly active editors for English Wikipedia in 2020\footnote{\url{https://stats.wikimedia.org/\#/en.wikipedia.org/contributing/active-editors/normal|line|2020-01-01~2021-01-01|(page_type)~content*non-content|monthly}} ), the labor cost of monitoring Wikipedia content quality and patrolling new edits (around 200k daily on English Wikipedia) is intensive. 
Automated strategies could be used to support the community of Wikipedia editors and reduce their workload, allowing editors to focus their efforts on more complex content moderation efforts. 
Despite recent advancements in the fields of Information Retrieval (IR) and Natural Language Processing (NLP), there exists only a few examples of successful automated support systems for Wikipedia content reliability monitoring, 
 the most well-known being ORES \cite{Halfaker20}, an open source service using multiple independent Machine Learning classifiers to score Wikipedia edits in real time. 

One of the main reasons for this gap is the lack of training data that researchers can use to detect and resolve content quality matters. To encourage research advancements on this front, in this paper we propose \textit{Wiki-Reliability}, a large dataset of English Wikipedia articles annotated with a wide range of  content reliability issues.

To create this dataset, we rely on Wikipedia's \textit{maintenance templates}, one of the key mechanisms for Wikipedia editors to monitor content quality. 
Templates appear as messages on articles, 
warning for quality issues within a page content (see Figure \ref{fig:teaser}). 
The usage of templates requires expert knowledge of Wikipedia community processes. Therefore, templates 
can be considered as expert-created labels, and 
the implicit data created as a by-product of Wikipedia editors workflow as reasonably high-quality crowd generated dataset.

\textit{Wiki-Reliability} focuses on the list of templates curated by the  {\verb|WikiProjectReliability|}\footnote{\url{https://en.wikipedia.org/wiki/Wikipedia:WikiProject_Reliability}} editors, who maintain templates related to citations and verifiability issues
, which are used to signal that moderation fixes are needed to improve article reliability . 

To create \textit{Wiki-Reliability}, we propose a novel methodology for template-based article labeling. This method can be easily reproduced to create new datasets using other types of templates on Wikipedia.
For each reliability template, we extract pairs of \textit{positive} and \textit{negative} versions of a Wikipedia article from its revision history. Positive examples are versions of an article which contain a reliability issue, signalled by the \textit{addition} of the template, while negative examples are article revisions where the issue has been resolved, signalled by the \textit{removal} of the template.  

Together with the positive/negative labels, for each sample in our dataset we also provide a set of metadata features, which help contextualize each data point, adding information that is not directly related with the textual content, for example the number of external links in an article, or the number of links pointing to other Wikipedia pages. We also parse the full article textual content, and include it as part of the dataset. We then identify a set of downstream research tasks enabled by the different template labels in our dataset. Finally, we test baselines for content reliability prediction, showing that the features in our data are informative, but that downstream tasks are not trivial to solve and require further research efforts. 


NLP and IR practitioners and researchers use Wikipedia extensively. This data will provide the opportunity for these communities to contribute back to Wikipedia, and help improve the quality of the online encyclopedia, while experimenting with large-scale data analysis techniques that can be generalized to other contexts across the World Wild Web. 
All the data and code generated for this project is being released with this paper.\footnote{\url{{https://github.com/kay-wong/Wiki-Reliability/}}} 
\section{Methodology for Data Labeling}
In this Section, we explain our proposed methodology to effectively extract high-quality annotated data from the unstructured  content of Wikipedia articles.

\subsection{Background: Wikipedia, Revisions, Templates and Wikiprojects}

Content on Wikipedia is dynamic. After an article is created, editors contribute by generating  updated versions. These versions are known as \textit{revisions}. Each revision has a numerical id, and is  stored in the editing history of the article. Revisions also come with associated metadata, such the timestamp, revisions' author and a comment added by the author summarizing the aim of their contribution. 

Wikipedia \textit{Templates}, also known as transclusions, are defined as ``pages created to be included in other pages''\footnote{https://en.wikipedia.org/wiki/Help:Template}. Editors use templates for a wide set of operations, such as creating infoboxes, navigational boxes, or creating warning messages. One of the most well known templates is the ``citation needed'' tag used to signal content requiring a citation. %

\textit{Wikiproject} is the concept used by Wikipedia editors to name groups of editors that collaborate on specific tasks. There are topic-focused Wikiprojects such \emph{WikiProject Science} or \emph{Wikiproject Sports}. But there are also groups of users that congregate to add and improve images and multimedia content to articles across topics, or to disambiguate page titles. In this paper we focus one group of users working on detecting and fixing problems related to content reliability. 
\setlength{\tabcolsep}{2.5pt}
\begin{table*}[!t]
\centering
{\small
\resizebox{\linewidth}{!}{
\begin{tabular}{l|ccll}
  \toprule
\textbf{Template} & \textbf{Count} & \textbf{Span}  & \textbf{Description}& \textbf{Downstream Tasks}\\
\midrule
\textit{Unreferenced} & 389966 & article & Article has no citations or references at all  & SR\\
\textit{One Source} & 25085 & article & Article cites only a single source.  & SR\\\hline
\textit{Original Research} & 19360 & article & Article contains original research. & CRP \\
\textit{More Citations Needed} & 13707 & article &  Article needs additional citations for verification.   & CRP, SR \\
\textit{Unreliable Sources} & 7147 & article & Some of the article's listed sources may not be reliable.  & CRP, SR\\
\textit{Disputed} & 6946 & article & Article's factual accuracy is disputed. & CRP \\
\textit{Pov} & 5214 & article/section & Article lacks a neutral point of view.  & CRP \\
\textit{Third-party} & 4952 & article & Article relies excessively on sources too closely associated with the subject  & CRP, SR \\
\textit{Contradict} & 2268 & article/section & Article or section contradicts itself  & CRP\\
\textit{Hoax} & 1398 & article & The truthfulness of an article is questioned, and is believed to constitute a hoax.  & CRP  \\
  \bottomrule
\end{tabular}
}}
\caption{List of revision templates in the \textit{Wiki-Reliability} dataset, together with the positive/negative pair counts, description, and ideas for potential downstream tasks ($CRP$= Content Reliability Prediction, $SR$= Source Retrieval).}
\label{table:template-counts}
\vspace{-2mm}
\end{table*}

\subsection{Selection of Templates}
\sloppy
To annotate Wikipedia article revisions, we start from the list of 41 templates curated by the {\verb|WikiProjectReliability|}. We categorize them according to their \textit{span}, namely the coverage of the template, i.e., if it indicates a reliability issue at \textit{article}, \textit{sentence} or \textit{section} level. We then manually sort the article-level maintenance templates based on their impact to Wikimedia, prioritizing templates which are of interest to the community.

\subsection{Parsing Wikipedia dumps}
Next, we search for the presence of templates in Wikipedia revisions. The full history of Wikipedia articles is available through periodically updated XML dumps\footnote{\url{https://dumps.wikimedia.org}}. We use the full English Wikipedia dump 
from September 2020, sizing 1.1\emph{TB} compressed in bz2 format. We convert this data to AVRO\footnote{\url{https://github.com/wikimedia/analytics-wikihadoop}} and process it using PySpark.  We apply a regular expression to extract all the templates in each revision, and retain all the articles that contain our predefined list of templates. Next, using the MediaWiki History dataset\cite{mediawiki_history} we obtain additional information (metadata) about each revision of the aforementioned articles. %

\subsection{Handling instances of vandalism}
The accurate detection of true positive and negative cases is further complicated by instances of vandalism, namely where a template has been maliciously/wrongly added or removed. To handle this, we rely on the wisdom of the crowd by ignoring revisions which have been reverted by other editors. 

Research by \cite{Kittur07} suggests that 94\% of reverts can be detected by matching MD5 checksums of revision content historically. %
However, comparing SHA checksums of consecutive revisions is a computationally expensive process as it requires processing through the entire history of revisions. Fortunately, the {\itshape MediaWiki History}\footnote{url{https://wikitech.wikimedia.org/wiki/Analytics/Data\_Lake/Edits/MediaWiki\_history}} table contains monthly data dumps of all events with pre-computed fields of computationally expensive features to facilitate analyses. We use the {\itshape revision\_is\_identity\_reverted} feature in revisions events, which marks whether a revision was reverted by another future revision, and exclude all reverted revisions from the articles' edit history.

\subsection{Obtaining positive and negative examples}
We next iterate through all consecutive revisions of articles' edit history to extract positive/negative class pairs for each template. 

We repeat the following process for all templates in our list. We retain as  \textit{positive} the first revision where a content-reliability template appears. Next, we look for \textit{negative} examples. Labeling negative samples in this context is a non-trivial task. While the presence of a template is a strong signal of a content reliability issue, the absence of a template does not necessarily imply the converse. A revision may contain a reliability issue that has not yet been reviewed by expert editors and flagged with a maintenance template. Therefore, we iterate through the article history succeeding the positive revision, and label as \textit{negative} the first revision where the template does not appear, i.e. the revision where the template was \textit{removed}. The negative example then acts as a contrasting class, as it constitutes the positive example which has been fixed for its reliability issue. 

This approach is particularly effective to ensure data coverage and quality.
While previous work selects positive and negative examples as pages \textit{with} and \textit{without} templates from a current snapshot of pages\cite{Anderka12, Bhosale13}, in our approach, we capture multiple instances of templates' additions and removals over the full history  of Wikipedia articles. By extracting pairs of positive/negative examples from the same article, we eliminate confounding factors such as the article topic, or its inherent quality, and provide data which is balanced across classes.


\subsection{Summary of the Data Labeling Pipeline}\label{sec:pipeline}
\begin{enumerate}
  \item {\verb|Obtain Wikipedia edit history|}:
  We download the Wikipedia dump available at September 2020~\footnote{\url{https://dumps.wikipedia.org}}, and load it in AVRO format for PySpark processing.

  \item {\verb|Obtain “reverted” status of a revision|}:
  We query the Mediawiki History dataset to extract the “reverted” status of a revision. 
  
  \item {\verb|Check if the revision contains a template|}:
  For each template in the template list, we loop over all non-reverted revisions to find the first revision where  the template has been added.   We mark such revision as \textit{positive} to indicate that the revision contains the template.
  
  \item {\verb|Process positive/negative pairs|}:
  
  We iterate through all consecutive revisions of a positive example to find the next non-reverted revision where the template has been removed, and mark it as \textit{negative}.

\end{enumerate}
\begin{table*}
\resizebox{\linewidth}{!}{
  \begin{tabular}{llc}
    \toprule
    \textbf{{Field}} & \textbf{{Description}}& \textbf{{Downstream Task}}\\
    \midrule
    \textit{page\_id} & Page ID of the revision & all\\
    \textit{revision\_id} & ID of the revision & all \\
    \textit{revision\_id.key} & ID of the corresponding pos/neg revision & all\\
    \hline
    \textit{txt\_pos} & Wikipedia's article in plain text when \textit{has\_template} is 1 & all\\
    \textit{txt\_neg} & Wikipedia's article in plain text when  \textit{has\_template} is 0 & all \\

    \hline

    \textit{revision\_text\_bytes} & Change in bytes of revision text & all\\
    \textit{stems\_length}  & Average length of stemmed text & all\\ 
    \textit{images\_in\_tags} & Count of images in tags & all \\
    \textit{infobox\_templates} & Count of infobox templates & all \\
    \textit{paragraphs\_without\_refs}  & Total length of paragraphs without references & all \\
    \textit{shortened\_footnote\_templates} & Number of shortened footnotes (i.e., citations with page numbers linking to the full citation for a source) & all \\
    \textit{words\_to\_watch\_matches} & Count of matches from Wikipedia's ``words to watch'': words that are flattering, vague or endorsing a viewpoint & all \\
    \textit{revision\_words} & Count of words for the revision & all\\
    \textit{revision\_chars} & Number of characters in the full article & all\\
    \textit{revision\_content\_chars} & Number of characters in the content section of an article & all  \\
    \textit{external\_links} & Count of external links not in Wikipedia & all\\
    \textit{headings\_by\_level(2)} & Count of level-2 headings & all\\
    \textit{ref\_tags} & Count of reference  tags, indicating the presence of a citation & all\\
  
    \textit{revision\_wikilinks} & Count of links to pages on Wikipedia & all \\
    \textit{article\_quality\_score} & Letter grade of article quality prediction & all \\
    \textit{cite\_templates} & Count of templates that come up on a citation link & CRP \\
    \textit{who\_templates} & Number of $who$ templates, signaling  vague ``authorities", i.e., "historians say", "some researchers" & SR \\
    \textit{revision\_templates} & Total count of all transcluded templates & SR\\
    \textit{category\_links} & Count of categories an article has  & SR \\    \hline

    \textbf{\textit{has\_template }(label)} & Binary label indicating presence or absence of a reliability template in our dataset & all\\
  \bottomrule
\end{tabular}}
  \caption{Schema of the \textit{Wiki-Reliability} dataset, with the fields extracted from the positive/negative revision examples in our data. Information for each template is stored in two files, one with the text and another with all the additional features.}
  \label{table:metadata-features}

\end{table*}

\section{Dataset Description}
We explain here the structure and post-processing of the final dataset released.
After replicating the process in Section \ref{sec:pipeline} for all the article-level templates, we select the templates with the largest amount of annotations (Section \ref{sec:reltemplates}),  extract the corresponding revision text, and compute metadata features (Sections \ref{sec:text} and \ref{sec:metadata-features}, respectively). The final dataset includes, for each template, the following information: page identifier, revision identifier, positive/negative template labels, text and features, as shown in Table \ref{table:metadata-features}.  We compiled two files for each template, one with annotations and metadata, and another one with the full revision text. 
The data is publicly available on Figshare\footnote{\url{https://figshare.com/articles/dataset/Wiki-Reliability_A_Large_Scale_Dataset_for_Content_Reliability_on_Wikipedia/14113799}}.

\subsection{Reliability Templates and Labels}\label{sec:reltemplates}
We repeat the positive/negative revision labeling for all the templates curated by {\verb|WikiProjectReliability|}, and assign, to each matching revision, a binary \textit{has\_template} label, which equals 0 for \textit{negative} revisions
, and 1 for \textit{positive} revisions. We filter out templates with positive/negative pair counts of less than 1000. The remaining top 10 templates are shown in Table~\ref{table:template-counts}.

\subsection{Downstream Tasks}
We also identify potential \textit{downstream tasks} that researchers in NLP and IR can perform on the data labeled with different templates, and assign them to each template in our dataset (see Table~\ref{table:template-counts}). 
We suggest two main tasks: \emph{i)} Content Reliability Prediction (CRP)  and \emph{ii)}  Source Retrieval (SR). The former consists in predicting the presence of a given template (the \emph{has\_template} column), while the latter consists in retrieving the sources and references that are relevant to the article and  improve the article quality. Note that although we provide a complete set of features plus the article content in plain text, additional information about references, links and categories can be extracted from the MediaWiki API\footnote{url{https://www.mediawiki.org/wiki/API:Tutorial}} using the page identifiers (see Table \ref{table:metadata-features}). 

For example, revisions labeled with templates such as \textit{Unreferenced} or \textit{One Source}, can be used as a real-world testbed by researchers interested in applications around reference retrieval or recommendation. The data labeled with templates like \textit{POV}, which indicates neutrality issues, or \textit{Hoax}, reflecting potential falsehoods in the article, can be used to train classifiers able to identify content reliability issues such as the presence of fake, partisan or malicious information, similar to previous work detecting the citation needed template~\cite{redi2019citation}.


\subsection{Metadata Features}\label{sec:metadata-features}
We extract metadata features for each positive/negative revision in our data by querying the ORES API’s Article Quality model\footnote{\url{https://www.mediawiki.org/wiki/ORES}}.  The model predicts the article quality class of a Wikipedia article according to the Wikipedia article quality grading scheme \footnote{\url{https://en.wikipedia.org/wiki/Template:Grading_scheme}}. The model generates features based on the structural characteristics of an article and its revision. Features can be obtained with a scoring request\footnote{\url{https://www.mediawiki.org/wiki/ORES/Feature_injection}}. We use the generated features along with the article quality score prediction as metadata features for our dataset, resulting in 26 metadata features in total.  We classified them according to the downstream task where they are used, removing features that would be artifact of the labels. 

\begin{figure*}[t]
\begin{subfigure}{0.5\linewidth}
    \includegraphics[width=\linewidth]{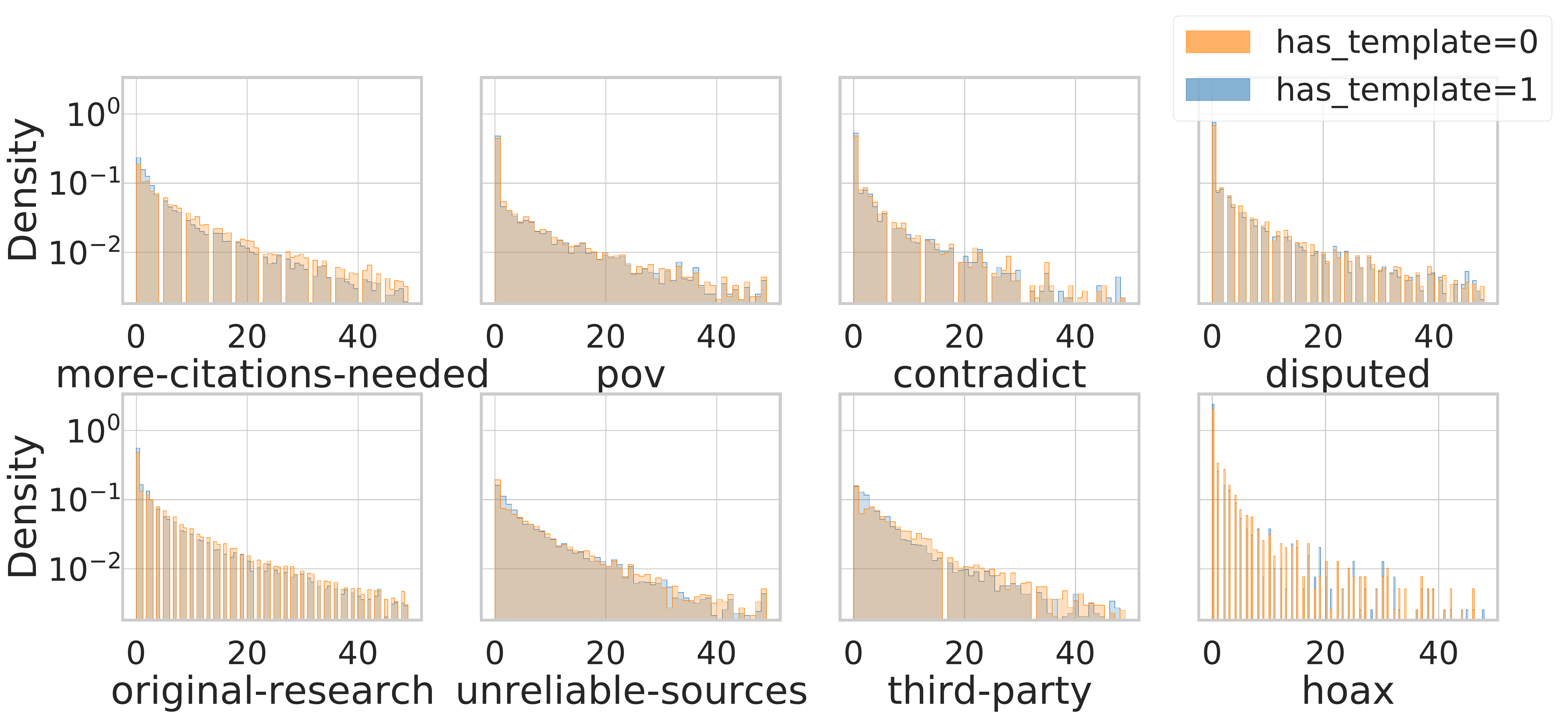}
    \caption{\textit{ref\_tags}.}
    \label{fig:featuredist_reftags}
\end{subfigure}%
\begin{subfigure}{0.5\linewidth}
    \includegraphics[width=\linewidth]{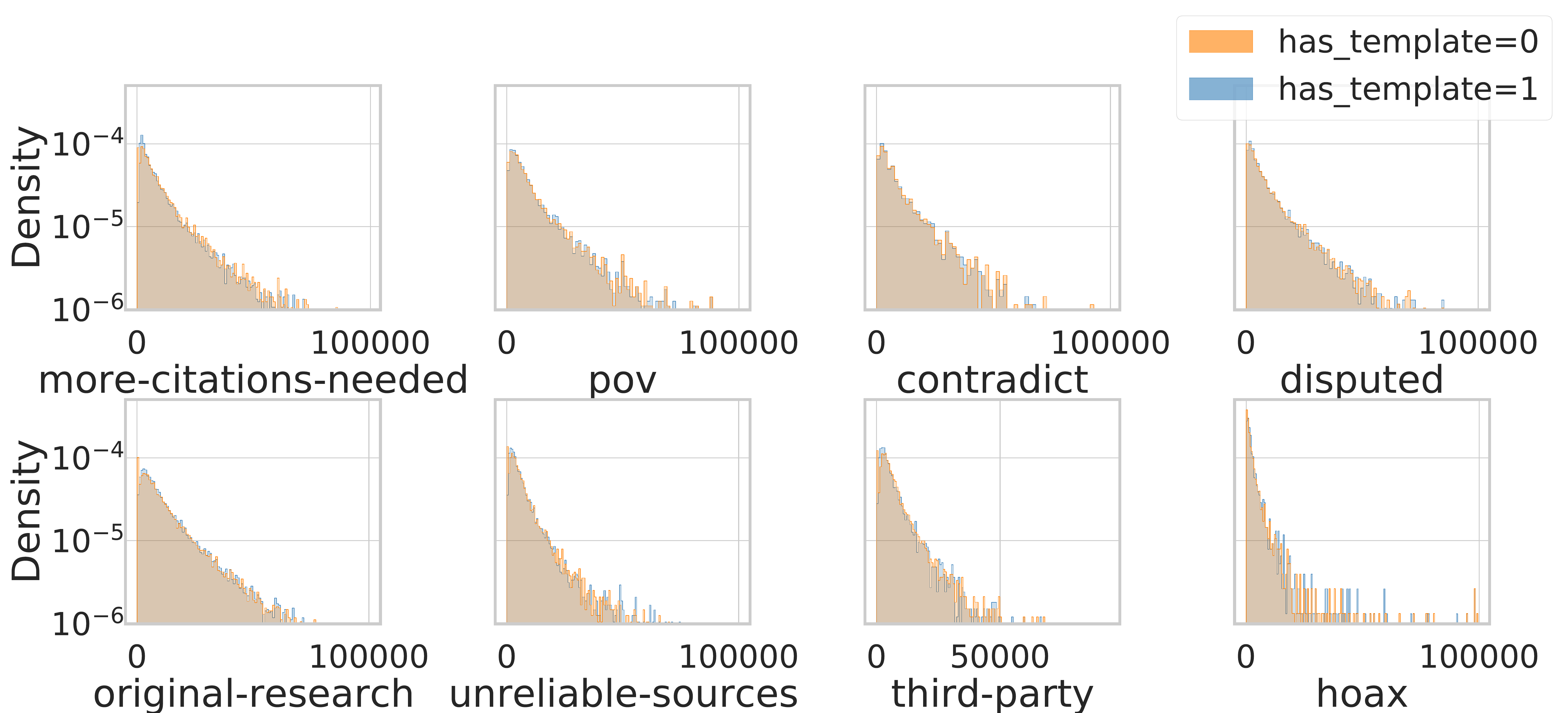}
    \caption{ \textit{revision\_chars}.}
    \label{fig:featuredist_paragraphswithoutrefs}
\end{subfigure}
\caption{Distribution of the \textit{ref\_tags} (a) and   \textit{revision\_chars} (b) features, reflecting respectively the number of reference tags in a revision and the length in characters of revision text, for positive and negative examples of each template.}
\label{fig:fig}
\end{figure*}
For our final dataset, we further narrow down the number of features to 20 , by selecting only the most important features\footnote{To analyze the importance of different features, we averaged the importance scores of features from different content reliability prediction models, and removed the ones with lower significance score. 
Note that the models trained on the new reduced subset of features achieves comparable (and sometimes improved) accuracy to the full set.} 
for template prediction tasks (see Section \ref{sec:Results} for more details).
The features released as part of the Wiki-Reliability dataset are listed in Table~\ref{table:metadata-features}.

Figures~\ref{fig:featuredist_reftags} and ~\ref{fig:featuredist_paragraphswithoutrefs} show the feature distribution for two features, {\itshape ref\_tags} and {\itshape revision\_chars}, across positive and negative revisions for different templates. We observe that, apart from \textit{Unreliable Sources} and \textit{Third Party}, articles revisions which have had their template removed, i.e respective reliability issue resolved, tend to have more reference tags,  suggesting that this metadata feature is actually informative of some reliability issues, as the feature distribution for positive revisions differ to the ones for negative revisions. Conversely, the number of characters in an article seems to be less informative of content reliability issues, since the \textit{revision\_chars} distribution tend to be similar across positive and negative examples for all templates.


\subsection{Content Text}\label{sec:text}
While some metadata features can be informative of the presence of a reliability template, researchers using this data might want to analyze the text of the revisions labeled as positive or negative. Models based on structural features might be unable to evaluate characteristics of the content such as writing quality, or the presence of tonal issues, which might be more essential for the prediction of templates such as {\textit{POV}} or \textit{contradiction}. Thus, we also create a text-based dataset for the purpose of text classification.

For each revision in our dataset, we query the API for its {\itshape wikitext}, which we parse using {\verb|mwparserfromhell|} to obtain only the plain text content, stripping out all wikilinks, templates, and tags. Further, we filter out all references sections, keeping only the main article content.

\section{Metadata-based Models and Analysis}


To better understand the content and the informativeness of our dataset, we design a set of classifiers for the downstream task of content reliability prediction. For each template where we identified $CRP$ as a potential downstream task (see Table \ref{table:template-counts}), we learn a model predicting the target label {\itshape has\_template} based on the metadata features. 

\subsection{Content Reliability Prediction Models}

We train content reliability prediction models using \textit{has\_template} as the target variable, and the  metadata features marked as $CRP$ as the independent variables. For each template, we fit 3 models: Logistic Regression, Random Forest, and Gradient Boosted Trees.

Each classifier is trained using 3-fold cross validation, with a train/test split ratio of 2:1, using the GroupKFold\footnote{\url{https://scikit-learn.org/stable/modules/generated/sklearn.model\_selection.GroupKFold.html}} iterator to enforce non-overlapping groups of articles across the training and test splits. This ensures that revisions of the same article will not appear in the test set if it already occurs in the training set, and vice versa, thus avoiding potential biases or overfitting. Training and test splits have balanced label distributions. 

For all experiments, we report the average classification accuracy, precision, recall and F1-score over the 3 folds of the cross-validation.






\subsection{Model Results}
\label{sec:Results}
\begin{figure}
    \centering
    \includegraphics[width=0.98\linewidth]{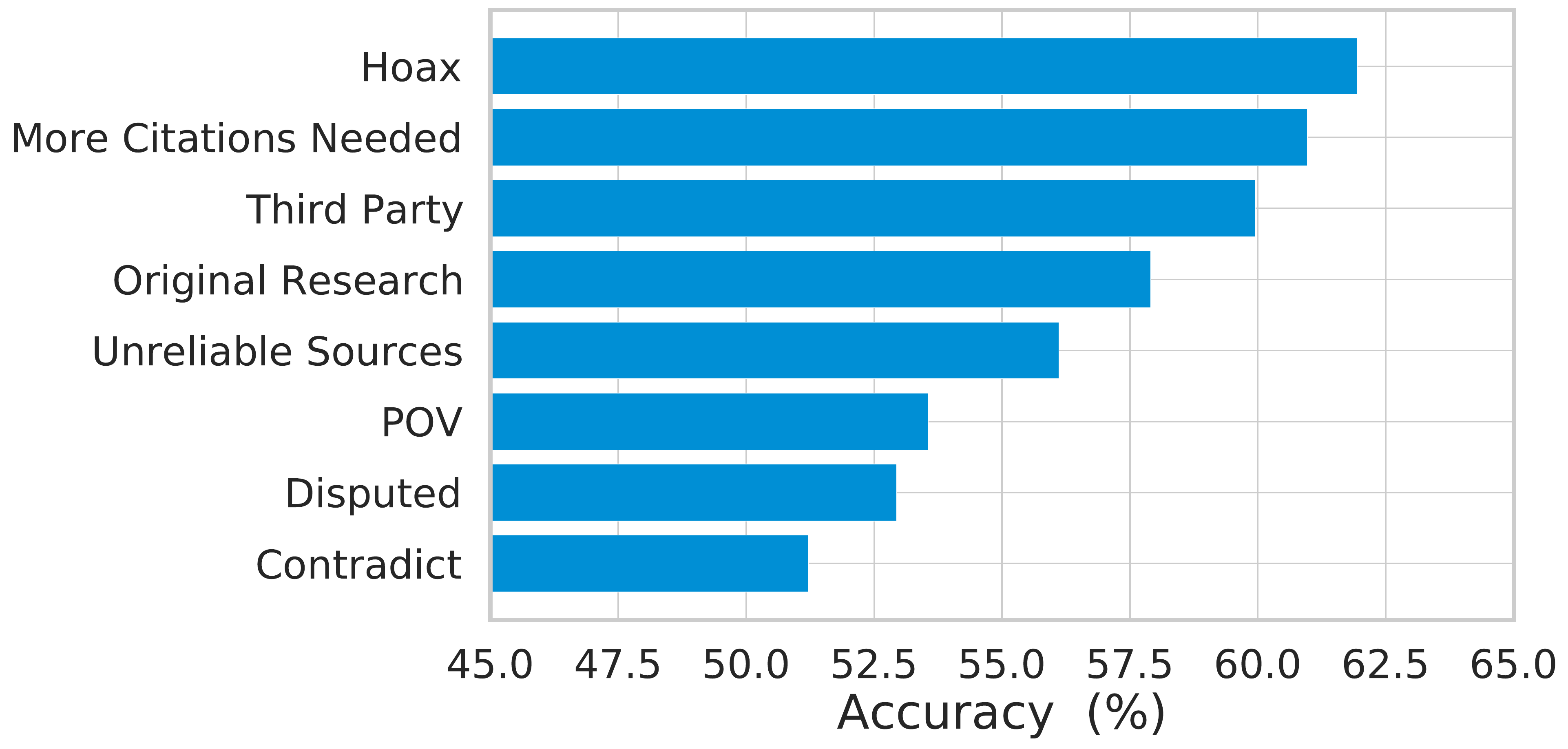}
    \caption{Average accuracy of the Gradient Boost-based prediction models for all predictable templates.}
    \label{fig:xgb-acc-results}
\end{figure}

Across all templates, the Gradient Boost model achieves the highest performance scores. Figure~\ref{fig:xgb-acc-results} shows the accuracy score results for all templates: models perform better than random, with the maximum accuracy for template prediction standing at 62\%. This suggests that the metadata features provided do carry some informativeness regarding the reliability of the articles' content. However, the relatively low performances of these models highlights the difficulty of this task, opening up opportunities for more research on this front.

Next, we look more in-depth at the top 3 templates by model accuracy: {\itshape Hoax}, {\itshape More Citations Needed} and {\itshape Third Party}. We present the full results of the classification  performances for all metrics and methods in Table ~\ref{table:results}. These results confirm that, across all different metrics, the Boosting-based models perform better than others, with the \textit{Hoax} template being more predictable than the other two.

\setlength{\tabcolsep}{1.5pt}
\begin{table}[t]
\resizebox{\linewidth}{!}{
\begin{tabular}{l c c c c c}
\toprule
\textbf{Template} & \textbf{Model} & \textbf{Accuracy} & \textbf{Precision} & \textbf{Recall} & \textbf{F1} \\
\midrule 

\textbf{Hoax} 
& LogReg & $0.58 \pm 0.033$ & $0.59 \pm 0.033$ & $0.58 \pm 0.033$  & $0.58 \pm 0.035$ \\
& RF & $0.58 \pm 0.015$  & $0.59 \pm 0.015$  & $0.58 \pm 0.015$  & $0.58 \pm 0.014$ \\
& \textbf{XGB} & $\textbf{0.62}  \pm 0.014$  & $0.62  \pm 0.012$  & $0.62  \pm 0.014$  & $0.62  \pm 0.015$ \\
\midrule 

\textbf{More Citations Needed} & LogReg & $0.56  \pm 0.002$  & $0.56  \pm 0.004$  & $0.56  \pm 0.003$  & $0.55  \pm 0.003$ \\
& RF & $0.58  \pm 0.005$  & $0.58  \pm 0.005$  & $0.58  \pm 0.005$  & $0.58  \pm 0.006$ \\
& \textbf{XGB} & $\textbf{0.61} \pm 0.000$  & $0.61 \pm 0.000$  & $0.610 \pm 0.000$  & $0.61 \pm 0.00$ \\

\midrule
\textbf{Third Party} & LogReg & $0.56  \pm 0.005$  & $0.58 \pm 0.005$  & $0.56 \pm 0.005$  & $0.56 \pm 0.006$ \\
& RF &  $0.57  \pm 0.006$  & $0.578 \pm 0.007$  & $0.57  \pm 0.006$  & $0.57  \pm 0.006$ \\
& \textbf{XGB} &  $\textbf{0.60} \pm 0.006$  & $0.600  \pm 0.005$  & $0.60 \pm 0.006$  & $0.59 \pm 0.006$ \\
\bottomrule
\end{tabular}
}
\caption{Metadata based models results for {\itshape Hoax}, {\itshape More Citations Needed} and {\itshape Third Party}  templates: comparison between Logistic Regression (\textit{LogReg}), Random Forests (\textit{RF}), and Gradient Boosted Trees (\textit{XGB})}
\label{table:results}
\vspace{-2mm}
\end{table}

\begin{figure}
    \centering
    \includegraphics[width=0.92\linewidth]{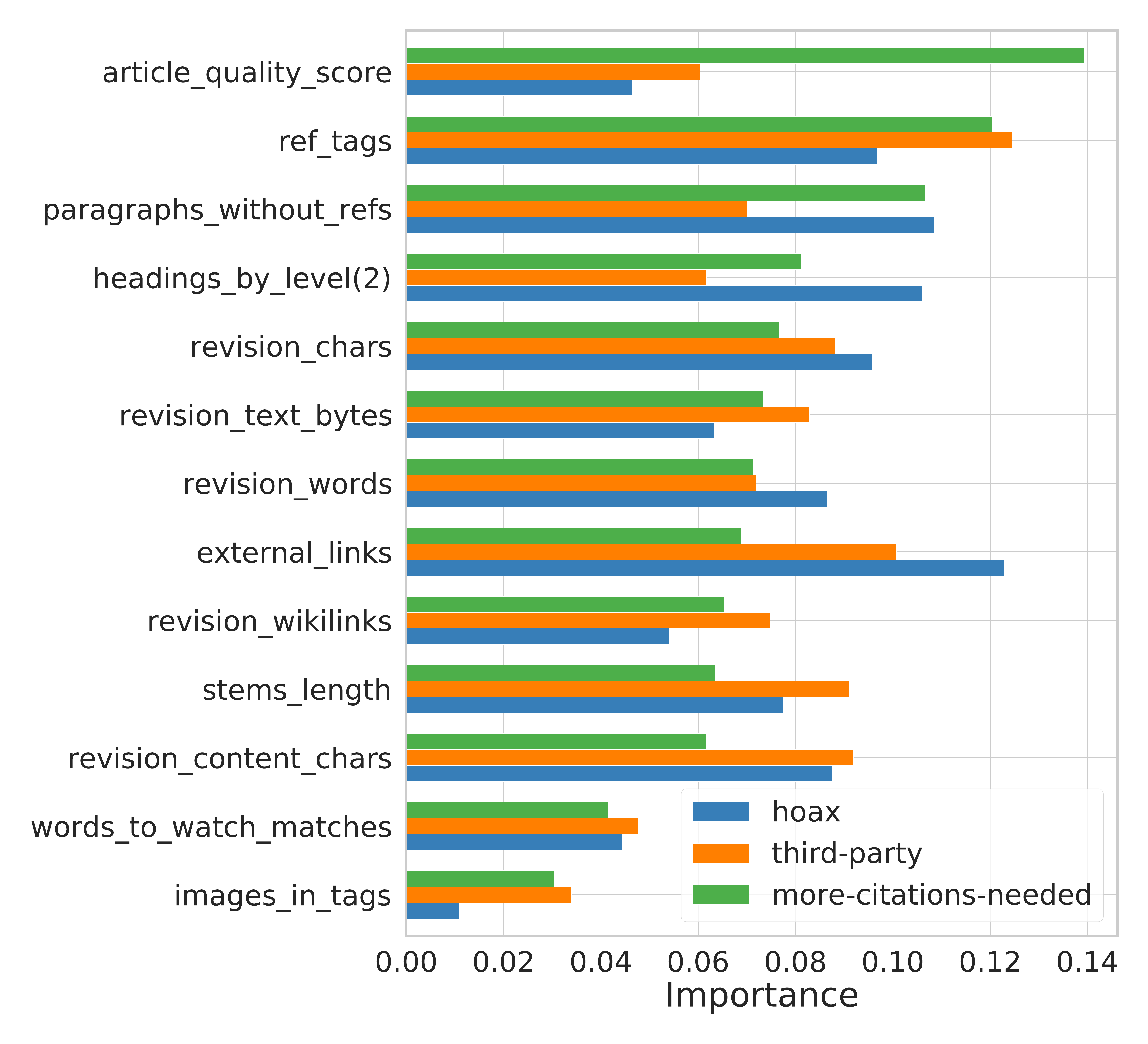}
    \caption{Feature importance for the models predicting the \textit{hoax}, \textit{more citations needed}, and \textit{third party} templates.}
    \label{fig:xgb-importance}
\end{figure}

To understand which features are most highly related with the presence of each template, we next obtain the feature importance scores computed from the Gradient Boosted Trees classifier. Feature importance is defined as the number of times a feature is used to split the data across all trees, divided by the total sum of occurrences. We compute these for the top 3 templates in Figure ~\ref{fig:xgb-importance}. 
We see that the presence of external links is the most significant feature for the \textit{hoax} template prediction: by looking at the feature distribution, we  see that the vast majority of articles tagged as ``hoax'' has indeed less than 2 external links, while pages where the ``hoax'' tags have been removed have a higher average number of external links. Article quality is a high predictor of the \textit{more citations needed} template, probably due to the fact that the article quality model incorporates features such as the \textit{ref\_tags} variable, which reflects the number of references in an article and tends to have much lower values in presence of reliability templates. Finally, across all three templates we see that the length of paragraphs without references is an indicator of reliability issues. While these results suggest that metadata-based models have potentials for content reliability prediction, future research will need to use language models trained on textual content to improve detection results. 
\section{Conclusion and Future Work}


    


The \textit{Wiki-Reliability} dataset contains Wikipedia articles 
labeled according to their reliability, using tags to signal specific reliability issues such as ``non-neutral point of view'', or ``disputed content''
.  We used the wisdom of experienced Wikipedia editors to label such articles, and additionally computed metadata features and release the articles' full text. \textit{Wiki-Reliability} is available for English Wikipedia only. Thanks to the reproduciblility of our labeling method,  we plan in the future to extend it to other languages.

We also suggested downstream tasks for this data. While these are just examples of what can be done with \textit{Wiki-Reliability}, we hope that our proposed method for dataset creation, together with the size and quality of the dataset generated, will foster creativity and promote research on novel NLP and IR tasks.  We showed that classifiers trained on metadata features perform reasonably well for some downstream tasks, thus demonstrating the inherent quality of the data and the predictability of Wikipedia templates. To verify these observations, a manual evaluation of the quality of data will be performed as part of our future work.

Though training complex language models was outside the scope of this work, 
a simple logistic regression model trained on TF-IDF features gave promising results for the $CRP$ task. This suggests that our data contains useful signals that a simple text-based model is able to capture, and we encourage researchers to use the large textual annotated data contained within this dataset to train more complex language models for content reliability tasks.  

With this dataset, we intend to provide tools and data for researchers to build effective tools for automated support of Wikipedia content moderation, thus allowing the scientific community to contribute back to Wikipedia, and help to improve the quality of the main repository of human knowledge. 
%
%
While the main focus of this work is Wikipedia, content reliability 
systems trained on this data could be designed to generalize to contexts outside of Wikipedia, such as news articles or social media posts. 



\bibliographystyle{ACM-Reference-Format}
\bibliography{bibliography}

%
\end{document}